\documentstyle[12pt,epsfig]{article}

\newcommand {\lan}    	{\langle}
\newcommand {\ran}    	{\rangle}

\newcommand {\ie}       {{\em i.e.}\ }
\newcommand {\overl}[1] {\overline{#1}}
\newcommand {\qm}	{\overline{q}}
\newcommand {\qea}	{q_{\rm EA}}
\newcommand {\hp}	{\tilde{h}}

\renewcommand {\d}    	{{\mbox d}}

\begin{document}

\title{On the Dynamics of the 4d Spin Glass in a magnetic field}

\author{
Giorgio Parisi, Federico Ricci-Tersenghi\\
and Juan J. Ruiz-Lorenzo\\[0.6em] 
{\small Dipartimento di Fisica and INFN, Universit\`a di Roma {\em La
Sapienza}}\\
{\small P. A. Moro 2, 00185 Roma (Italy)}\\[0.3em]
{\small \tt giorgio.parisi@roma1.infn.it}\\
{\small \tt riccife@chimera.roma1.infn.it}\\
{\small \tt ruiz@chimera.roma1.infn.it}\\[0.5em]
}

\date{November 13, 1997}

\maketitle

\begin{abstract}

We study the four dimensional Gaussian spin glass in presence of a
magnetic field. Using off-equilibrium numerical simulations we have
found that the probability distribution of the overlaps is built in
the same way as that of the Mean Field approximation with replica
symmetry breaking. Finally we have studied the violation of the
fluctuation-dissipation theorem in presence of magnetic field.

\end{abstract}  

\thispagestyle{empty}
\newpage

\section{\protect\label{S_INT}Introduction}

One of the main open questions in the study of finite dimensional spin
glasses is the existence of a phase transition in presence of an
external field.  While in absence of magnetic field the droplet
model~\cite{DROPLET} and the Mean Field (MF) picture~\cite{MEPAVI}
both predict the existence of a phase transition, in presence of
magnetic field the situation is fully different: Mean Field predicts a
phase transition whereas the droplet model shows that the magnetic
field destroys the frozen phase.

Moreover there are some analytical arguments which imply that the
phase transition in presence of magnetic field is of a rather peculiar
type.  For instance Bray and Roberts~\cite{BRAY_ROB}, working with a
reduced theory (obtained by projecting the original theory in the
replicon subspace; in presence of magnetic field that is the only
critical mode), have shown that there is no weak-coupling fixed point
in magnetic field in their renormalization group equations also near 6
dimensions.  An absence of a weak-coupling fixed point is often taken
as an indication of a first order transition.  Here the situation is
quite less clear.

The very existence of a transition is still controversial and not too
much work has been devoted to its study.  Numerical simulations have
been done in the past \cite{EQUIL}.  They were compatible with the
possibility of a transition, but the situation was not so clearly cut
and no convincing conclusions could be reached.  Only recently an
off-equilibrium numerical simulations in 4 dimensions have strongly
suggested the existence of a phase transition in presence of magnetic
field~\cite{MAPAZU}.

Two main advantages in using dynamical methods are that we can
simulate very large systems (up to $40^4$ in this work) loosing
practically all the finite size effects and that they are quicker with
respect to an equilibrium simulation, because we don't need to
thermalize.  These methods have been already largely used in the
numerical studies of spin glasses (see for instance
reference~\cite{BOOK}).

In the broken replica symmetry solution of the SK
model~\cite{MEPAVI,parisibook2} there is a phase transition also when
the system is plunged into a magnetic field and the line $T_{\rm
c}(h)$ which separates the paramagnetic from the spin glass phase is
called Almeida-Thouless (AT) line~\cite{AT}.  The order parameter of
the Mean Field theory is the probability distribution of the overlaps,
$P(q)$.

In absence of external field $P(q)$ is a delta function centered on
$q=0$ for $T>T_{\rm c}$, while for $T<T_{\rm c}$ it becomes a highly
non trivial function with two delta functions on the values $q=\pm
\qea$ ($\qea$ is the maximum allowed value for the overlap) and a
non-zero part between them.

When a magnetic field is switched on the function $P(q)$ becomes zero
for every negative overlap and the minimum allowed value for $q$ is
shifted upwards from $-q_{\rm EA}$ to $q_{\rm min}>0$, while the
maximum value ($\qea$) almost doesn't change.  This means, in terms of
the distribution function of the overlaps, that for $T>T_{\rm c}(h)$
the function $P(q)$ is a delta function centered on a strictly
positive value and that for $T<T_{\rm c}(h)$ the function $P(q)$ is
the sum of a delta function on the maximum value $q_{\rm max} = \qea$
plus a non-zero part down to $q_{\rm min}>0$ and a smaller weighted
delta function on $q_{\rm min}$.

In this paper we present evidences for a Mean Field like phase
transition at finite temperature.  We show that the order parameter
$P(q)$ has a non-zero support: we have characterized numerically the
mean, the minimum and the maximum value allowed, denoted $\qm, q_{\rm
min}$ and $q_{\rm max}$ respectively, and we have found that $q_{\rm
min} < \qm < q_{\rm max}$.

The plan of the paper is the following.  In the next section we fix
the notation and we describe the quantities we have measured.  In
sections three we show the numerical results.  Finally we present the
conclusions.

\section{\protect\label{S_OBS}The model and the observables}

We have simulated the Gaussian Ising spin glass in four dimensions on
a hypercubic lattice of volume $N=L^4$ with periodic boundary
conditions.  The Hamiltonian of the system is given by
\begin{equation}
{\cal H}=-\sum_{<ij>}\sigma_i J_{ij}\sigma_j -\sum_i\sigma_i h_i\ .
\protect\label{ham}
\end{equation}
By $<ij>$ we denote the sum over nearest neighbor pairs.  The $J_{ij}$
are Gaussian variables with zero mean and unitary variance.  The
external field is also Gaussian with zero mean and variance $h_0^2$.
We have studied systems with $h_0=0.05,0.1,0.2,0.3,0.5$.

We can justify the choice of a Gaussian magnetic field as follows. The
starting point is the Ising spin glass Hamiltonian with an uniform
magnetic field $h_0$
\begin{equation}
{\cal H}_0=-\sum_{<ij>}\sigma_i J_{ij}\sigma_j - h_0\sum_i\sigma_i\ .
\end{equation}
We can perform the following transformation (a ``local gauge
transformation'') on the couplings: $J_{ij} \to J^\prime_{ij} \equiv
n_i J_{ij} n_j$, where $n_i=+1$ or $-1$. This transformation leaves
the Hamiltonian ${\cal H}_0$ unchanged because the probability
distribution of the couplings is Gaussian
($J_{ij}^2=(J^\prime_{ij})^2$). Now we recast the spins to $s_i\equiv
n_i \sigma_i$, and finally our Hamiltonian reads
\begin{equation}
 {\cal H}_0=-\sum_{<ij>}s_i J_{ij}s_j - \sum_i (h_0 n^\prime_i) s_i\ ,
\end{equation}
where $n^\prime_i n_i=1$. We remark the full arbitrariness of the
choice of $n^\prime_i$'s. In particular we can choose them from a
bimodal distribution: i.e. $n^\prime_i=1$ with probability $1/2$ and
$-1$ with the same probability.  And so, if we define $h^\prime_i
\equiv h_0 n^\prime_i$, the Hamiltonian can be written as
\begin{equation}
{\cal H}_0=-\sum_{<ij>}s_i J_{ij}s_j - \sum_i h^\prime_i s_i\ .
\end{equation}
We have therefore shown that a spin glass with an uniform magnetic
field ($h_0$) is equivalent to a spin glass in which the magnetic
field is random with zero mean and variance $h_0^2$.  The probability
distribution of such a magnetic field is bimodal, not Gaussian.
Nevertheless there are not reasons to suppose a different physical
behavior of these two cases (bimodal and Gaussian).  We have chosen a
Gaussian distribution, and not a bimodal one, because in the Gaussian
case there are exact relations among some quantities.

We are interested in measuring the mean value of the overlap between
two replicas, which is defined as $\qm = \int q P(q)\d q$, without
doing expensive equilibrium simulations.  If we take two replicas in
random configurations (as we do at the beggining of a simulation)
their overlap is zero ($q(t\!=\!0)\!=\!0$).  Letting them evolve, the
overlap will never increase beyond $q_{\rm min}$, defined as the
minimum overlap allowed at the equilibrium~\footnote{This can be
better understood if we think that $P(q)$ is the equilibrium
distribution of a dynamical variable which feels a potential
$V(q)=-\log P(q)$.  The peak at $q=q_{\rm min}$ in $P(q)$ corresponds
to a deep well in $V(q)$ and if $q$ starts with a value less than
$q_{\rm min}$ it will go downhill towards $q_{\rm min}$.  See
reference \cite{MAPARU} for more details.}.  This fact has been
largely verified also in off-equilibrium simulations without magnetic
field: during the simulation the overlap fluctuates around zero or
slightly grows.  This observation gives us a practical tool to
calculate $q_{\rm min}$ via an infinite-time extrapolation, but also
asserts that we cannot get information on the whole $P(q)$ simply by
looking at the off-equilibrium overlap.

To measure $\qm$ we have exploited a relation valid at equilibrium
when the applied field is Gaussian, which reads
\begin{equation}
\frac{\overl{\lan \sigma_i h_i\ran}}{h_0^2} = \frac{1-\qm}{T} \ ,
\protect\label{rel}
\end{equation}
where with the overline we mean an average over the quenched random
interactions and external fields.  This relation can be easily
obtained via an integration by parts (exploiting that $h_i$ is a
Gaussian random variable) and it's exact also in a finite volume.  In
order to compute $\qm$ we can measure $\overl{\lan \sigma_i(t)
h_i\ran}$, which is a quantity that rapidly converges to its
infinite-time value.  The fact that $q_{\rm min}$ differs from $\qm$
is a clear signal of replica symmetry breaking.

The second part of our study is focused on the fluctuation-dissipation
theorem (FDT) and its generalization in the out of equilibrium
regime~\cite{OUT,FRARIE,FDT}, called off-equilibrium
fluctuation-dissipation relation (OFDR).  In reference~\cite{FDT} a
detailed study of such a relation in finite dimensional spin glasses
without magnetic field can be found.  Here we extend those studies in
presence of a magnetic field, obtaining similar results, which confirm
the Mean Field behavior of the phase transition.

To study the OFDR we have measured the spin-spin autocorrelation
function, $C(t,t_w)$, and the integrated response of the system,
$\overl{\lan \sigma_i \hp_i \ran}/\epsilon_0^2$, where the
perturbation to the Hamiltonian $\cal H$ (eq.(\ref{ham})), $\hp$, is a
random Gaussian magnetic field with zero mean and variance
$\epsilon_0^2$.  In the next paragraphs we will obtain a formula that
links, even in the early times of the dynamics, the response and the
auto-correlation function.

Given a quantity $A(t)$ that depends on the local variables of our
original Hamiltonian ($\cal H$).  We can define the associate
autocorrelation function
\begin{equation}
C(t,t^\prime) \equiv \lan A(t) A(t^\prime) \ran \ ,
\label{auto}
\end{equation}
and the response function
\begin{equation}
R(t,t^\prime) \equiv \left. \frac{\delta \lan A(t) \ran}{\delta
\epsilon(t^\prime)}\right|_{\epsilon=0} \ ,
\label{res}
\end{equation}
where we have assumed that the original Hamiltonian has been perturbed
by a term
\begin{equation}
{\cal H}^\prime= {\cal H} + \int \epsilon(t) A(t)\,\d t \ .
\end{equation}
The brackets $\lan (\cdot\cdot\cdot) \ran$ in eq.(\ref{auto}) and
eq.(\ref{res}) imply here a double average, one over the dynamical
process and a second over the disorder.

In the dynamical framework assuming time translational invariance it
is possible to derive the fluctuation-dissipation theorem, that reads
\begin{equation}
R(t,t^\prime)=\beta \theta(t-t^\prime) \frac{\partial
C(t,t^\prime)}{\partial t^\prime} \ .
\protect\label{FDT}
\end{equation}
In spin models a common choice for $A(t)$ is $A(t) = \sigma_i(t)$ or
$A(t) = N^{-1/2} \sum_i \sigma_i(t)$.  In this case, because the
system feels a magnetic field, to have a simpler response we should
perturb it with a random field, $\hp$, and measure the staggered
magnetization.  In order to derive a fluctuation theorem where the
response is related to the one site correlation, we must chose a
perturbation such that the off-site elements of the response are zero
\footnote{We notice also that in absence of a magnetic fields, a
constant field perturbation is gauge equivalent to a random magnetic
field.}. So here we put $A(t) = N^{-1/2} \sum_i n_i \sigma_i(t)$,
where $n_i = \hp_i / \epsilon_0$.  Thanks to the fact that $\lan n_i
n_j \ran = \delta_{i,j}$, we have that with this choice
\begin{equation}
C(t,t^\prime) = \frac1N \sum_i \lan \sigma_i(t)\sigma_i(t^\prime)\ran
\protect\label{corr}
\end{equation}
and
\begin{equation}
R(t,t^\prime) = \frac{\delta m_{\rm s}[\hp](t)}{\delta \hp(t^\prime)}
\ , \protect\label{resp}
\end{equation}
where
\begin{equation}
m_{\rm s}[\hp](t) = \frac1N \sum_i \lan n_i \sigma_i(t) \ran
\protect\label{stag}
\end{equation}
is the staggered magnetization, which is a functional of the magnetic
field, $\hp(t)$, and a function of the time.

The fluctuation-dissipation theorem holds in the equilibrium regime,
but in the early times of the dynamics we expect a breakdown of its
validity.  Mean Field studies \cite{CUKU} suggest the following
modification of the FDT:
\begin{equation}
R(t,t^\prime)=\beta X(t,t^\prime) \theta(t-t^\prime) \frac{\partial
C(t,t^\prime)}{\partial t^\prime} \ .
\end{equation}
It has also been suggested in \cite{CUKU,FM,BCKP} that the function
$X(t,t^\prime)$ is only a function of the autocorrelation:
$X(t,t^\prime)=X(C(t,t^\prime))$.  We can then write the following
generalization of FDT, which should hold in early times of the
dynamics, the off-equilibrium fluctuation-dissipation relation (OFDR),
that reads
\begin{equation}
R(t,t^\prime)=\beta X(C(t,t^\prime)) \theta(t-t^\prime) \frac{\partial
C(t,t^\prime)}{\partial t^\prime}\ .
\protect\label{OFDR}
\end{equation}
We can use the previous formula, eq.(\ref{OFDR}), to relate the
observable quantities defined in eq.(\ref{corr}) and eq.(\ref{stag}).
Using the functional Taylor expansion we can write
\begin{equation}
m_{\rm s}[\hp](t) = m_{\rm s}[0](t) + \int_{-\infty}^\infty \d
t^\prime ~ \left.\frac{\delta m_{\rm s}[\hp](t)}{\delta \hp(t^\prime)}
\right|_{\hp(t)=0} \hp(t^\prime) + {\rm O}(\hp^2) \ .
\end{equation}
Exploiting the definition of eq.(\ref{resp}) and using the fact that
$m[0](t)=0$ for every perturbing field orthogonal to the pre-existing
one, \ie such that $\sum_i \lan n_i h_i \ran = 0$ (which is true if
$h_i$ is another (uncorrelated) random field, as happens in our case),
we obtain
\begin{equation}
m_{\rm s}[\hp](t) = \int_{-\infty}^t \d t^\prime ~ R(t,t^\prime)
\hp(t^\prime) + {\rm O}(\hp^2) \ .
\end{equation}
This is just the linear-response theorem neglecting higher orders in
$\hp$.

By applying the OFDR we obtain the dependence of the staggered
magnetization with time in a generic time-dependent magnetic field
(with a small strength), $\hp(t)$,\footnote{The symbol $\simeq$ means
that the equation is valid in the region where linear-response holds.}
\begin{equation}
m_{\rm s}[\hp](t) \simeq \beta \int_{-\infty}^t \d t^\prime ~
X[C(t,t^\prime)] \frac{\partial C(t,t^\prime)}{\partial t^\prime}
\hp(t^\prime) \ .
\end{equation}
Now we can perform the following experiment.  We let the system evolve
with the unperturbed Hamiltonian of eq.(\ref{ham}) from $t=0$ to
$t=t_w$, and then we turn on the perturbing magnetic field $\hp$,
which is Gaussian distributed with zero mean and time-independent
variance, $\epsilon_0^2$.  Finally, with this choice of the magnetic
field, we can write\footnote{We ignore in our notation the fact that
$m_{\rm s}[\hp](t)$ depends also on $t_w$.}
\begin{equation}
m_{\rm s}[\hp](t) \simeq \epsilon_0 \beta \int_{t_w}^t \d t^\prime ~
X[C(t,t^\prime)] \frac{\partial C(t,t^\prime)}{\partial t^\prime} \ ,
\protect\label{mag_1}
\end{equation}
and by performing the change of variables $u=C(t,t^\prime)$, equation
(\ref{mag_1}) reads
\begin{equation}
m_{\rm s}[\hp](t) \simeq \epsilon_0 \beta \int_{C(t,t_w)}^1 \d u ~
X[u] \ ,
\protect\label{mag_2}
\end{equation}
where we have used the fact that $C(t,t) \equiv 1$ (always true for
Ising spins).  In the equilibrium regime (FDT holds, $X=1$) we must
obtain
\begin{equation}
m_{\rm s}[\hp](t) \simeq \epsilon_0 \beta (1 - C(t,t_w)) \ ,
\protect\label{mag_fdt}
\end{equation}
\ie $m_{\rm s}[\hp](t)\, T/\epsilon_0$ is a linear function of
$C(t,t_w)$ with slope --1.  We remark that we can use this formula to
obtain $q_{\rm max}$ as the point where the curve $m_{\rm s}[\hp](t)$
versus $C(t,t_w)$ leaves the line with slope $-\beta \epsilon_0$ (as
we will explain).

In the limit $t, t_w \to \infty$ with $C(t,t_w) = q$, one has that
$X(C) \to x(q)$, where $x(q)$ is given by
\begin{equation}
x(q)=\int_{q_{\rm min}}^q \d q^\prime ~P(q^\prime)\ , 
\protect\label{x_q}
\end{equation}
where $P(q)$ is the equilibrium probability distribution of the
overlap. Obviously $x(q)$ is equal to 1 for all $q > q_{\rm max}$, and
we recover FDT for $C(t,t_w) > q_{\rm max}$.  This link between the
dynamical function $X(C)$ and the static one $x(q)$ has been already
verified for finite dimensional spin glasses~\cite{FDT}.

For future convenience, we define 
\begin{equation}
S(C)\equiv \int_C^1 \d q ~x(q) = \int_C^1 \d q ~\int_{q_{\rm min}}^q
\d q^\prime ~ P(q^\prime)\ .
\protect\label{s_c}
\end{equation}
or equivalently
\begin{equation}
P(q) = -\left.\frac{\d^2 S(C)}{\d^2 C}\right|_{C=q} \ .
\label{pq}
\end{equation}  
In the limit where $X \to x$ we can write eq.(\ref{mag_2}) as
\begin{equation}
\frac{m_{\rm s}[\hp](t)\;T}{\epsilon_0} \simeq S(C(t,t_w)) \ .
\protect\label{final}
\end{equation}

Looking at the relation between the correlation function and the
integrated response function for large $t_w$ we can thus obtain
$q_{\rm max}$, the maximum overlap with non-zero $P(q)$, as the point
where the function $S(C)$ becomes different from the function $1-C$,
and $q_{\rm min}$ as the smallest value of $C$.

At this point we have numerical methods to compute three important
different values of $q$: $q_{\rm max}, q_{\min}$ and $\qm$.

\section{\protect\label{S_NUM}Numerical results}

\subsection{\protect\label{S_NUM_A}$\qm$ and $q_{\rm min}$}

We are interested in the behavior of the system in the out of
equilibrium regime, so we do not need to thermalize the sample and we
can simulate very large samples of millions of spins ($24^4$, $32^4$
and $40^4$). We expect our data not to be affected by large finite
size bias: we find that, in the range of temperature considered, the
data for different lattice sizes ($L=24,32,40$) coincide within the
error-bars, with the largest systems ($L=32,40$) giving practically
the same values.

All the numerical simulations have been performed on the parallel
super-computer APE100~\cite{APE}.

In the first part of our study we have done simulations using the
Hamiltonian of eq.(\ref{ham}) without perturbing the system ($\hp=0$).

According to our dynamical approach, we are interested in doing
measurements in the off-equilibrium regime of large times.  So we
perform the simulations following an annealing schedule, with slower
and slower cooling rates.  This is equivalent to run a simulation at a
fixed temperature, considering larger and larger waiting
times~\cite{FDT,4DIM}.  The advantage is twice: within a single run we
are able to collect data at different temperatures and there are
smaller finite-time effects because when the temperature is lowered by
an amount $\Delta T$ the time needed to ``forget'' the previous
temperature is smaller than if it had been quenched from $T=\infty$.
In the present case the range of temperatures~\footnote{We remind that
in the four dimensional Gaussian Ising spin glass without external
field the transition temperature is $T_{\rm c}=1.8$~\cite{4DIM}.} is
from $T=3.0$ down to $T=0.5$ with a step of $\Delta T=0.25$.  The
cooling rates have been chosen in such a way that in the $s$-th
annealing run the number of Monte Carlo Steps (MCS) is proportional to
$2^s$, with $s$ ranging from 0 to 12.  The annealing procedure is the
following: indexing the temperatures from the highest ($i_{T=3.0}=1$)
to the lowest ($i_{T=0.5}=11$) the number of MCS at each temperature
is $2^s \cdot i_T^2$.  This means that in the slowest run the system
will stay at the lowest temperature for about half a million MCS.

To analyze the data, we fix a temperature, we recollect the data at
that temperature from different annealing runs and we extrapolate the
result in the limit of infinitely slow cooling.  This will give us
information on the large times off-equilibrium regime.

\begin{figure}
\begin{center}
\leavevmode
\centering\epsfig{file=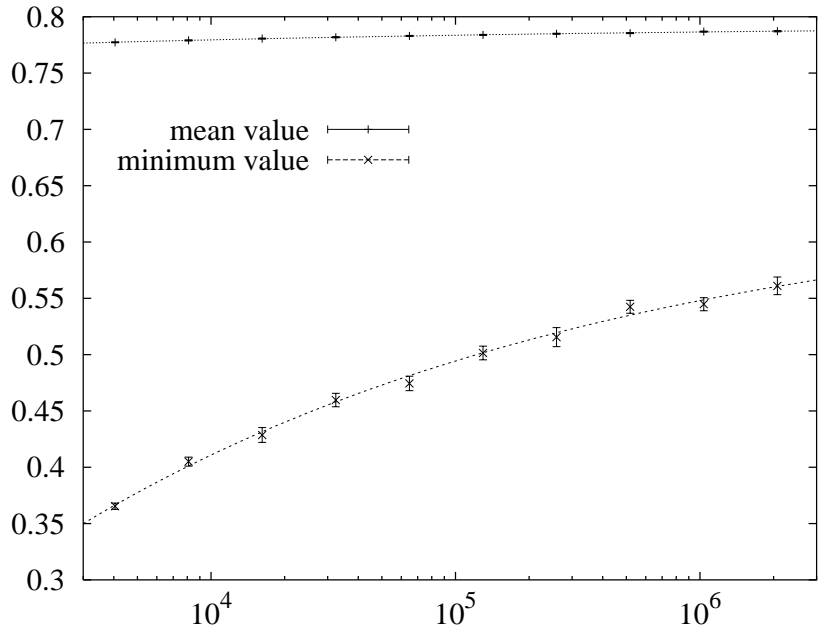,width=0.75\linewidth}
\end{center}
\protect\caption{Data of $q(t)$ (bottom) and $1 - \overl{\lan
\sigma_i(t) h_i \ran}\; T / h_0^2$ (top) versus Monte Carlo time. The
curves are the best fits (see text). $L=40$, $h_0=0.3$ and $T=0.5$.}
\protect\label{F_extr}
\end{figure}

In particular we look at the overlap between two replicas, $q(t)$,
which start very far from each other: $q(t=0)=0$.  In the limit
$t\to\infty$ this overlap will tend to the minimum overlap allowed at
the equilibrium, $q_{\rm min}$.  Also we have measured $\overl{\lan
\sigma_i(t) h_i \ran}$ from which we obtain the value of $\qm$ using
eq.(\ref{rel}).  In figure~\ref{F_extr} we show the data ($L=40$,
$h_0=0.3$ and $T=0.5$) whose infinite-time extrapolation gives the
mean and the minimum value for the overlap (top and bottom data
respectively).

We fit the data plotted in figure~\ref{F_extr} with the following
formula: $q(t) = A t^{-B} + C$.  If the value of the exponent $B$ were
too small the uncertainty on the value of $C$ should be very large and
the significance of the fit very poor.  We have found that it
decreases with decreasing temperature or magnetic field and we have
checked in all our fits that it were not too small.  In
figure~\ref{F_extr} we present the worst case we can fit satisfactory
(lowest temperature and $h_0=0.3$), obtaining as the best parameters:
$A=-0.069(2)$, $B=0.20(3)$ and $C=\qm=0.792(5)$ for the top data and
$A=-1.4(2)$, $B=0.20(3)$ and $C=q_{\rm min}=0.64(2)$ for the bottom
data.  Both fits have very good $\chi^2$ values.  Note that the best
$B$ exponent is the same in both fits.  The presence of a power law
approach to equilibrium can be associated to the existence of a well
defined off-equilibrium correlation length in finite dimensional spin
glasses which grows with a power law of the
time~\cite{4DIM,MAPARURI,6DIM}.

The dynamical approach in presence of a magnetic field has the
advantage that, using a fitting function of the type $A t^{-B} + C$
for any observable in the off-equilibrium regime, the best value for
the exponent $B$ is always greater than in the case of exactly zero
external field and moreover it has a finite limit when $T\to 0$, while
for $h=0$ very often $B\propto T$.  The reasons for this difference in
behaviour are unclear to us~\footnote{We would like to recall that
numerical simulations of relative small size systems for the SK model
indicate that the time for passing at equilibrium from one value of
$q$ to an other value of $q$ with the same sign increases as $\exp(A
N^{1/4})$ while that for changing the sign of $q$ increases as $\exp(A
N^{1/2})$.  In magnetic field only the first processes are present and
in the SK model they correspond to much faster movements.}.

Now we clearly see the usefulness of the trick of using a Gaussian
magnetic field: we can calculate $\qm$ from the extrapolation of
$\overline{\lan \sigma_i(t) h_i \ran}$, which is a practically
constant quantity with respect to $q(t)$ (look at
figure~\ref{F_extr}).  Moreover we don't need to do any limit of small
$h$ and so we don't have the problem of very small $B$ exponents.

\begin{figure}
\begin{center}
\leavevmode \centering\epsfig{file=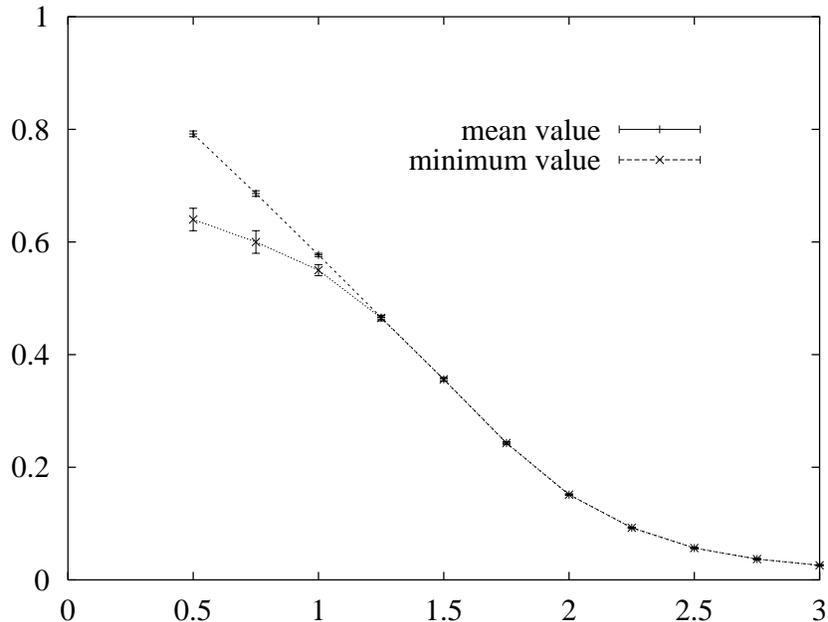,width=0.75\linewidth}
\end{center}
\protect\caption{$\qm$ (top) and $q_{\rm min}$ (bottom) versus $T$ for
$L=40$ and $h_0=0.3$.}
\protect\label{F_q_T}
\end{figure}

The same kind of fitting analysis has been done for all temperatures,
obtaining values for $q_{\rm min}(T)$ and $\qm(T)$, plotted in
figure~\ref{F_q_T}.  In the high temperature phase the extrapolated
values for $q_{\rm min}$ and $\qm$ coincide (as they should because
the equilibrium overlap distribution function is a delta function)
confirming the correctness of the dynamical method.

Figure~\ref{F_q_T} gives clear evidence of a wide region ($T < T_{\rm
c}(h_0=0.3) \simeq 1.2$) where the order parameter $P(q)$ is not a
single delta function.  We have calculated the functions $q_{\rm
min}(T)$ and $\qm(T)$ also for different values of the magnetic field.
We see a clear bifurcation also for $h_0=0.5$ at a temperature $T_{\rm
c}(h_0=0.5) \simeq 1.0$, while for lower magnetic fields the results
are less clean because the errors are larger (due to the just
described problem in the extrapolation procedure).

\begin{figure}
\begin{center}
\leavevmode \centering\epsfig{file=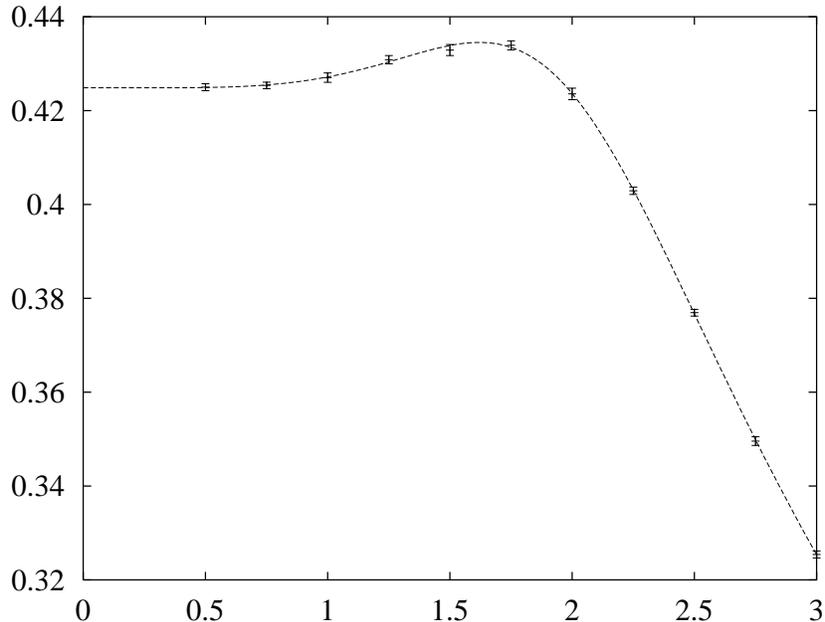,width=0.75\linewidth}
\end{center}
\protect\caption{$\overl{\lan \sigma_i h_i \ran}/ h_0^2$ versus $T$
for $L=40$ and $h_0=0.3$. The curve is only a guide to the eye.}
\protect\label{F_mT}
\end{figure}

The last result we present in this section is the shape of the
equilibrium magnetization in a magnetic field as a function of the
temperature.  More precisely what we show in fig~\ref{F_mT} is the
staggered magnetization $\overl{\lan \sigma_i h_i \ran}/ h_0^2$, but
we have verified that it behaves qualitatively and also almost
quantitatively like the magnetization in a constant field.

We observe a behavior similar to the one measured in experiments on
real samples: the magnetization has a little peak, whose height is
greater than the zero-temperature value by a few percents.  We want to
stress that the peak is at a temperature higher than the critical one,
so the magnetization peak is not exactly on the AT-line.

\subsection{\protect\label{S_NUM_B}Fluctuation-Dissipation in presence
of magnetic field}

In the second part of our study we have simulated an Ising spin glass
in a Gaussian magnetic field with variance $h_0^2$ ($h_0=0.2,0.3,0.5$)
at a fixed temperature ($T=0.75,1.0,1.5$).  We have measured, for
various waiting times ($t_w=2^8,2^{11},2^{14},2^{17}$), the
autocorrelation function, defined in eq.(\ref{corr}), and the
staggered magnetization, defined in eq.(\ref{stag}), with different
amplitudes of the perturbing field ($\epsilon_0=0.02,0.03,0.06$).  We
are interested in the relation between these two quantities, which
tends to the function $X(C)$ in the large $t_w$ limit (see
eq.(\ref{final})).  The data do not depend on the choice of
$\epsilon_0$ or $L$, so the results we will show are in the
linear-response regime and they are not affected by large finite size
biases.  In these runs the starting configuration of the spins is
always random.

We can keep in mind the possible link between the static and dynamics
(that we do not study in the present paper) for a spin glass in
presence of a magnetic field and we can compute the qualitative form
of the function $X(C)$ using as input the $S(C)$ function obtained in
the Mean Field approximation and in the droplet model.

From the function $S(C)$ we can get information on the overlap
distribution function $P(q)$, through eq.(\ref{pq}).  Let us remind
which is the prediction for the $S(C)$ assuming the validity of one of
the competing theories described in the introduction.  The droplet
model predicts $P(q)=\delta(q-{\hat q})$ and consequently
\begin{equation}
S(C) = \left\{
\begin{array}{cl}
1 - {\hat q} & {\rm for} \;\; C \le {\hat q} \ ,\\
1 - C & {\rm for} \;\; C > {\hat q} \ ,
\end{array}
\right.
\end{equation}
\ie there is no dependence of the staggered magnetization on
$C(t,t_w)$ in the off-equilibrium regime ($C \le {\hat q}$), like an
ordered ferromagnet~\cite{BARRAT}.  On the other hand the MF like
prediction for the overlap distribution~\cite{MEPAVI} $P(q) = (1-x_M)
\delta(q-q_{\rm max}) + x_m \delta(q-q_{\rm min}) + \tilde{p}(q)$
(where the support of $\tilde{p}(q)$ belongs to the interval $[q_{\rm
min},q_{\rm max}]$), implies that
\begin{equation}
S(C) = \left\{
\begin{array}{cl}
S(0) & {\rm for} \;\; C \le q_{\rm min} \ ,\\
\tilde{s}(C) & {\rm for} \;\; q_{\rm min} < C \le q_{\rm max} \ ,\\
1 - C & {\rm for} \;\; C > q_{\rm max} \ ,
\end{array}
\right.
\end{equation}
where $\tilde{s}(C)$ is a quite smooth and monotonically decreasing
function such that
\begin{equation}
\tilde{p}(q) = -\left.\frac{\d^2\tilde{s}(C)}{\d C^2}\right|_{C=q}\ .
\end{equation}

\begin{figure}
\begin{center}
\leavevmode \centering\epsfig{file=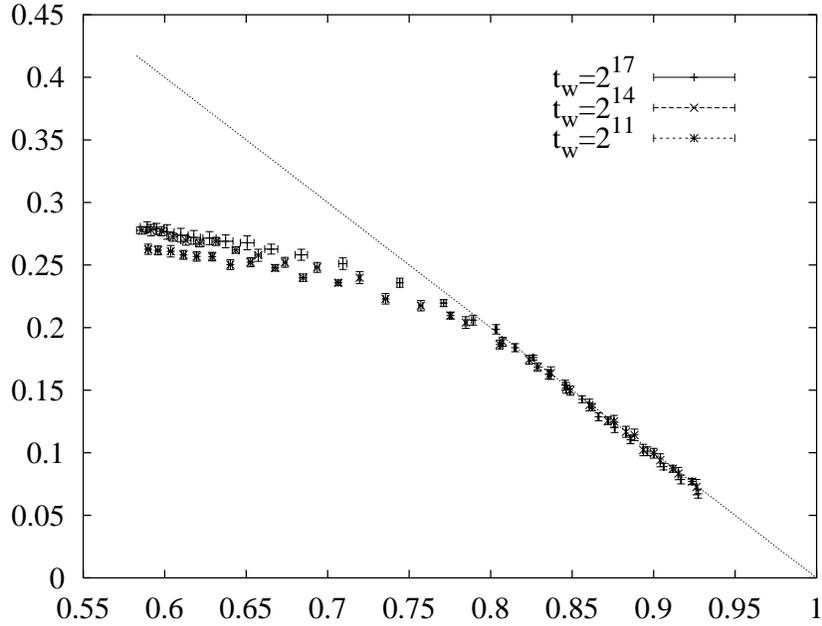,width=0.75\linewidth}
\end{center}
\protect\caption{$m_{\rm s}[\hp](t)\;T/\epsilon_0$ versus $C(t,t_w)$,
with $T=0.75$, $h_0=0.3$ and $\epsilon_0=0.06$. The straight line is
$1-C$.}
\protect\label{F_fdt_low}
\end{figure}

In figure~\ref{F_fdt_low} we plot $m_{\rm s}[\hp](t)\;T/\epsilon_0$
versus $C(t,t_w)$, with $T=0.75$, $h_0=0.3$ and $\epsilon_0=0.06$.
The data are the average over 6 samples of a $32^4$ system.  We can
see from figure~\ref{F_fdt_low} that, when the data are not on the
straight line (FDT or quasi-equilibrium regime), they do not lie on an
horizontal line, \ie they depend on the value of the autocorrelation
also in the off-equilibrium regime.  So we can conclude that the
droplet theory is not able to describe the data in the frozen phase.

Interpreting the data using the MF picture, we deduce from the small
curvature of the function $S(C)$ in the region $C \le q_{\rm
max}$~\footnote{The data can be fitted also with a straight line,
whose non-zero inclination is almost $t_w$-independent.} that
$\tilde{p}(q)$ is small.  Anyway the existence of two different delta
functions in $P(q)$ is clear.

\begin{figure}
\begin{center}
\leavevmode \centering\epsfig{file=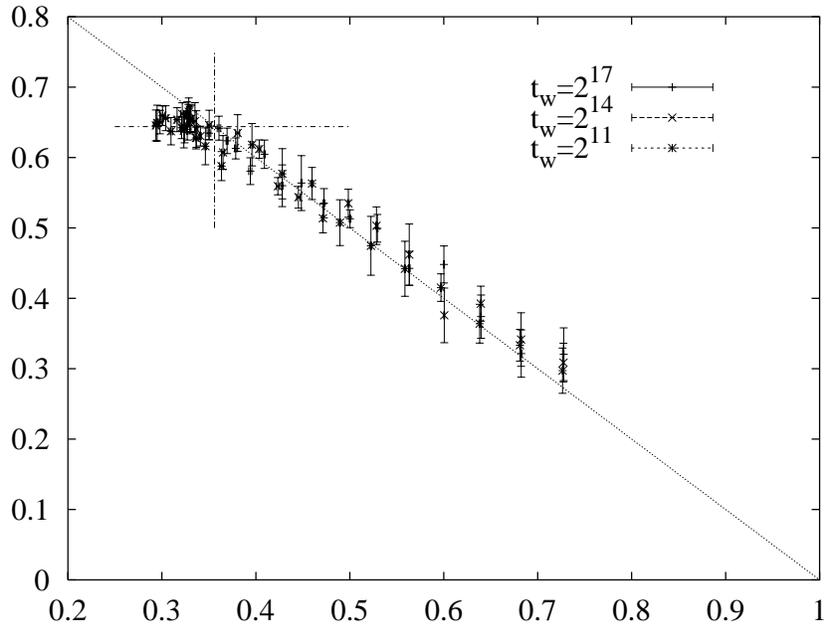,width=0.75\linewidth}
\end{center}
\protect\caption{$m_{\rm s}[\hp](t)\;T/\epsilon_0$ versus $C(t,t_w)$,
with $T=1.5$, $h_0=0.3$ and $\epsilon_0=0.02$. The straight line is
$1-C$. See text for the short vertical and horizontal lines.}
\protect\label{F_fdt_high}
\end{figure}

To verify the correctness of the method we present in
figure~\ref{F_fdt_high} the same kind of plot for a system which is in
the paramagnetic phase ($T=1.5$, $h_0=0.3$ and $\epsilon_0=0.02$).
The data stay, as they should, on the equilibrium line $S(C)=1-C$.
The few points which leave the straight line are those with the lowest
$t_w$ and large $t$.  This effect is due to the fact that the
autocorrelation $C(t,t_w)$ tends to $q_{\rm min}$ only if the system
at time $t_w$ has reached equilibrium.  On the contrary if the system
is in a random configuration at time $t_w$ the autocorrelation will
tend to zero also in a magnetic field.  In our simulation the system
is in an intermediate situation and so the lowest value for the
autocorrelation with $t_w=2^{11}$ is something smaller than $q_{\rm
min}$, while for $t_w=2^{17}$ we think that the system is nearly
equilibrated and the autocorrelation does not decrease beyond $q_{\rm
min}$.  In figure~\ref{F_fdt_high} we have reported the value for the
equilibrium overlap, $\qm = q_{\rm min} = q_{\rm max} \simeq 0.356$,
calculated with the annealing runs (vertical line) and the
corresponding staggered magnetization (horizontal line).  Though it is
not very clear from figure~\ref{F_fdt_high}, we have verified that for
every $t_w$ the staggered magnetization saturates to the value marked
with an horizontal line and that for the greater waiting time the
autocorrelation tends to the vertical line.

\section{\protect\label{S_CONCLUSIONS}Conclusions}

We have obtained, using off-equilibrium simulations, $q_{\rm min},
q_{\rm max}$ and $\qm$ for the four dimensional Gaussian spin glass in
presence of a magnetic field finding that in the low temperature
region $q_{\rm min} <\qm<q_{\rm max}$ according with the predictions
of Mean Field Theory.

This result point clearly toward to a phase transition between a spin
glass phase with spontaneously broken replica symmetry ($q_{\rm min}
<\qm<q_{\rm max}$) and a phase where the replica symmetry is stable
($q_{\rm min}=\qm=q_{\rm max}$).

Moreover we have extended the numerical studies of the validity of the
fluctuation-dissipation theory obtaining that the function that
determines the violation depends only on the correlation, as Mean
Field theory predicts. We plan in the future to try to link this
function ($X$) with that obtained from the static of the system ($x$)
as it has been done in absence of magnetic field although some
indirect evidences of this link has been shown in the present paper.

\section{Acknowledgments}

We acknowledge interesting discussions with E. Marinari and
F. Zuliani.  J.~J.~Ruiz-Lorenzo is supported by an EC HMC
(ERBFMBICT950429) grant.


\newpage

\begin{thebibliography}{100}

\bibitem{DROPLET} W. L. McMillan, J. Phys. C {\bf 17}, 3179
(1984). A. J. Bray and M. A. Moore, J. Phys. C {\bf 18}, L699
(1985). G. J. Koper and H. J. Hilhorst, J. de Physique {\bf 49}, 429
(1988). D. S. Fischer and D. A. Huse, Phys. Rev. B {\bf 38}, 386
(1988); Phys. Rev. B {\bf 38}, 373 (1988).

\bibitem{MEPAVI} M. M\'ezard, G. Parisi and M. A. Virasoro, {\em Spin
Glass Theory and Beyond}. World Scientific (Singapore 1987).

\bibitem{BRAY_ROB} A. J. Bray and S. A. Roberts, J. Phys. C {\bf 13},
5405 (1980).

\bibitem{EQUIL} S. Caracciolo, G.Parisi, S. Patarnello and N. Sourlas,
Europhys. Lett. {\bf 11}, 783 (1990); J. de Physique {\bf 51}, 1877
(1990). E. R. Grannan and R. E. Hetzel, Phys. Rev. Lett. {\bf 67}, 907
(1991). J. C. Ciria, G. Parisi, F. Ritort and J. J. Ruiz-Lorenzo,
J. de Physique I {\bf 3}, 2207 (1993). M. Picco and F. Ritort, J. de
Physique I {\bf 4}, 1619 (1994); cond-mat/9702041.

\bibitem{MAPAZU} E. Marinari, G. Parisi and F. Zuliani,
cond-mat/9703253.

\bibitem{BOOK} E. Marinari, G. Parisi and
J. J. Ruiz-Lorenzo. ``Numerical Simulations of Spin Glass Systems'' in
``Spin Glasses and Random Fields", edited by P. Young. World
Scientific (Singapore 1997). cond-mat/9701016.

\bibitem{parisibook2} G.Parisi, {\em Field Theory, Disorder and
Simulations}. World Scientific (Singapore 1992).

\bibitem{AT} J. R. L. de Almeida and D. J. Thouless, J. Phys. A:
Math. Gen. {\bf 11}, 983 (1978).

\bibitem{MAPARU} E. Marinari, G. Parisi and J. J. Ruiz-Lorenzo. In
preparation.

\bibitem{OUT} J.-P. Bouchaud, L. F. Cugliandolo, J. Kurchan,
M. M\'ezard. ``Out of equilibrium dynamics in spin-glasses and other
glassy systems'' in ``Spin Glasses and Random Fields", edited by
P. Young. Word Scientific (Singapore 1997).  cond-mat/9702070

\bibitem{FRARIE} S. Franz and H. Rieger, J. Stat. Phys. {\bf 79} 749
(1995).

\bibitem{FDT} E. Marinari, G. Parisi, F. Ricci-Tersenghi and
J. J. Ruiz-Lorenzo, cond-mat/9710120.

\bibitem{CUKU} L. F. Cugliandolo and J. Kurchan, Phys. Rev. Lett.
{\bf 71}, 173 (1993); Philosophical Magazine {\bf 71}, 501 (1995); J.
Phys. A: Math. Gen. {\bf 27}, 5749 (1994).

\bibitem{FM} S. Franz and M. M\'ezard, Europhys. Lett. {\bf 26}, 209
(1994).

\bibitem{BCKP} A. Baldassarri, L. F. Cugliandolo, J. Kurchan and
G. Parisi, J. Phys. A: Math. Gen. {\bf 28}, 1831 (1995).

\bibitem{APE} C. Battista {\em et al.}, Int. J. High Speed Comp. {\bf
5}, 637 (1993).

\bibitem{4DIM} G. Parisi, F. Ricci-Tersenghi and J. J. Ruiz-Lorenzo,
J. Phys. A: Math. Gen. {\bf 29}, 7943 (1996).

\bibitem{MAPARURI} E. Marinari, G. Parisi, J. J. Ruiz-Lorenzo and
F. Ritort. Phys. Rev. Lett {\bf 76}, 843 (1996).

\bibitem{6DIM} G. Parisi, P. Ranieri, F. Ricci-Tersenghi and
J. J. Ruiz-Lorenzo, J. Phys. A: Math. Gen. {\bf 30}, 7115 (1997).

\bibitem{BARRAT} A. Barrat, cond-mat/9710069.

\end{thebibliography}
\end{document}